\documentclass{article}
 \pdfoutput=1
\usepackage{graphicx}  % standard LaTeX graphics tool;
\usepackage{amsmath}   % For getting proper math eqns
\usepackage{amssymb}   % Used for getting various symbols eg. gtrsim, lesssim etc.
\usepackage[export]{adjustbox}
\usepackage{bm} % For bold math (esp of lower greek letters)
\usepackage{dcolumn}% Align table columns on decimal point
\usepackage{color}
\usepackage{mathrsfs}
\usepackage{amsfonts}
\usepackage{varioref}

\usepackage{wrapfig}

\RequirePackage[colorlinks,citecolor=blue,urlcolor=magenta,linkcolor=blue]{hyperref}

\labelformat{section}{Section #1} 
\labelformat{subsection}{Section #1} 
\labelformat{subsubsection}{Section #1}
\labelformat{subsubsubsection}{Section #1}
\labelformat{equation}{Eq.~(#1)} 
\labelformat{figure}{Fig.~#1} 
\labelformat{subfigure}{Fig.~\thefigure#1} 
\labelformat{table}{Tab.~#1} 
\labelformat{appendix}{Appendix #1}
\title{\bf Comments on the entropic gravity proposal}
\author{\bf Sourav Bhattacharya$^{1}$\footnote{sbhatta@iitrpr.ac.in}~, ~Panagiotis Charalambous$^{2}$\footnote{pchara21@ucy.ac.cy}~, ~Theodore N Tomaras$^{3}$\footnote{tomaras@physics.uoc.gr}~,\\
\bf ~~and~Nicolaos Toumbas$^{2}$\footnote{nick@ucy.ac.cy}\\
$^{1}$\small{Department of Physics, Indian Institute of Technology Ropar, Rupnagar, Punjab 140 001, India}\\
$^{2}$\small{Department of Physics, University of Cyprus, Nicosia 1678, Cyprus}\\
$^{3}$\small{ITCP and Department of Physics, University of Crete, 700 13 Heraklion, Greece} }

\begin{document}
  
\maketitle
\begin{abstract}
\noindent
Explicit tests are presented of the conjectured entropic origin of the gravitational force. The gravitational force on a test particle in the vicinity of the horizon of a large Schwarzschild black hole in arbitrary spacetime dimensions  is obtained as entropic force.  The same conclusion can be reached for the cases of a large electrically charged black hole and a large slowly rotating Kerr black hole. The  generalization along the same lines to a test mass in the field of an arbitrary spherical star is also studied and found not to be possible. Our results thus reinforce the argument that the entropic gravity proposal cannot account for the gravitational force in generic situations.
\end{abstract}
\vskip .5cm 
\noindent
%{\bf PACS : } \\
\noindent
{\bf keywords : Emergent gravity, entropic force, alternative gravity, holography} 
\bigskip
%%%%%%%%%%%%%%%%%%%%%%%%%%%%%%%%%%%%%%%%%%%%%%%%%%%%%%%%%%%%%%%%%%%%%%%%%%%%%%%%%%
\section{Introduction}\label{s1}
%%%%%%%%%%%%%%%%%%%%%%%%%%%%%%%%%%%%%%%%%%%%%%%%%%
\noindent
Motivated by black hole physics \cite{Beke,Hawking,Beke2}, the holographic principle \cite{tHooft,Susskind} and string theoretic developments on emergent space
and the AdS/CFT correspondence \cite{Maldacena}, Verlinde argues that gravity should be understood as an entropic effect caused by
the tendency of the underlying microscopic theory to maximize entropy \cite{Verlinde,Verlinde2}. Other pertinent work includes \cite{Jacobson, takayanagi, VanRaamsdonk, Padmanabhan, Li, Cai, Lashkari, Susskind2, Kiritsis:1999ke}.
The purpose of this paper is to demonstrate some concrete tests
of the conjecture in various spacetime dimensions.

In particular, we revisit the problem of a particle freely falling in the vicinity of the horizon of a large $D$-dimensional Schwarzschild
black hole. As is well known, a large black hole
is in a state of (near) thermal equilibrium, with entropy given by (one quarter of) the horizon area in Planck units. 
The static observer outside the black hole experiences
a thermal environment with increasing temperature as we move toward the horizon. 
Placing a particle at some distance away from the horizon perturbs the geometry and causes the horizon area to change.
In the thermodynamical picture,
thermal equilibrium is disturbed and the entropy acquires dependence on the distance of the particle from the horizon. 
We investigate whether the gravitational force on the particle (as seen by the static observer) can be interpreted as an entropic force.
Once the particle gets absorbed by the black hole, thermal equilibrium is restored and the entropy of the system becomes maximal.

In the limit of large black hole mass, the near horizon region becomes sufficiently 
weakly curved so as to obtain the backreaction on the geometry due to the test particle.
More specifically when the particle is slowly moving,
we obtain the shift of the horizon area $\delta A$, the entropy change $\delta S$ and the
temperature $T$ in terms of the particle's distance $\rho$ from the horizon. Using the formula
\begin{equation}
F=T \frac{dS}{d\rho}
\end{equation}
we compute the entropic force and find that it agrees with the gravitational force on the particle
{\it irrespectively of the number of spacetime dimensions}.
The same conclusion is reached next for the case of a  charged test particle in the vicinity of  the horizon of a large charged black hole, as well as for a test particle in the near horizon region of a slowly rotating large Kerr black hole. To the best of our knowledge, no explicit calculations exist in the literature, demonstrating that the gravitational force on a test particle in the field of a black hole of arbitrary mass (angular momentum and/or other charges) can be obtained as an entropic force. In this work we present such a computation in the limit of large black hole mass and test particle near the horizon.

We then proceed to investigate the case of a test mass moving in the field of an arbitrary
spherically symmetric mass distribution, not necessarily a black hole. The static Schwarzschild observer is
locally equivalent to a Rindler observer, uniformly accelerating with respect to an inertial frame. However, in this case our approximations cannot be used to obtain the backreaction of the test mass on the Rindler horizon. 
Instead, we use a holographic spherical screen sufficiently close to the observer
and associate to it an entropy and a temperature. We find that the entropy shift needed to interpret the gravitational force on a nearby test particle as an entropic force is precisely given by one quarter of the change of the screen area in Planck units. The shift in the area arises due to the backreaction of the test particle on the geometry. We argue that this result and the assumption of  thermal equilibrium imply that the entropy on the screen already saturates the holographic bound. This in turn requires the mass distribution to have collapsed to a black hole and the screen to be the black hole horizon. It seems difficult to modify some of the underlying assumptions in order to realize the entropic scenario in the more generic cases. {\it The only explicit examples for which the gravitational force can be obtained as an entropic force involve a slowly moving test particle in the near horizon region of a large black hole.}

Our arguments reinforce other objections concerning the entropic gravity proposal, such as the irreversibility effects of entropic forces and possible inconsistencies with the interference patterns in ultracold neutron experiments~\cite{Kobakhidze:2010mn, Nesvizhevsky}. For a possible inconsistency of the entropic gravity proposal with that of MOND~\cite{Milgrom}, we refer our reader to~\cite{Dai:2017qkz}.

We also refer our reader to discussions on the so called emergent gravity paradigm pertaining to the thermodynamic aspects of gravity~\cite{Padmanabhan:2009vy, Padmanabhan:2002sha}, where the thermodynamic characteristics of the Einstein equations in various contexts have been discussed. See also~\cite{Chakraborty:2015hna, Chakraborty:2015aja, Chakraborty:2014rga} for recent applications of this formalism to null hypersurfaces which are not necessarily Killing horizons, to higher derivative alternative gravity theories and also in cosmology.

The paper is organized as follows. In \ref{s2} we consider radial motion of a test particle in the gravitational
field of a Schwarzschild black hole in $D$ dimensions. We obtain the backreaction on the geometry in the limit of large black hole mass
and the corrected thermodynamic quantities at the moment the particle is instantaneously at rest.  We proceed to compute the entropic
force and find agreement with the gravitational force for all $D \ge 4$. Some results on the metric perturbation are reviewed in \ref{A1}.
The $D$-dimensional mean value theorem -- reviewed in \ref{A2} -- plays a crucial role in the computation. We also consider the case of a large Reissner-Nordstrom black hole and the effect of
spacetime rotation by considering the Kerr metric.
In \ref{s3} we analyze the more general spherically symmetric case and argue that the entropic gravity proposal cannot account for the gravitational force in generic situations. Finally our results, perspectives and open
problems are summarized in the discussion \ref{s4}.

%%%%%%%%%%%%%%%%%%%%%%%%%%%%%%%%%%%%%%%%%%%%%%%%%%%%%%%%%%%%%%%%%%%%%%%%%%%%%%%%%%
\section{Test particle in the gravitational field of a Schwarzschild black hole in D dimensions}\label{s2}
%%%%%%%%%%%%%%%%%%%%%%%%%%%%%%%%%%%%%%%%%%%%%%%%%%
\noindent

We are interested in the motion of a test particle of mass $m$ in the near horizon region of a Schwarzschild black hole in $D=d+1$ spacetime dimensions ($D\ge 4$).
The metric is given by
\begin{equation}
ds^2 = -f\left( r \right)dt^2 + \frac{dr^2}{f\left( r \right)} + r^2d\Omega_{D-2}^2,\,\,\,\,\,f(r)=1-\left(R_S\over r\right)^{D-3}\label{SchwarzschildMetric}
\end{equation}
where
\begin{equation}
R_S = \left(\frac{16\pi GM}{(D-2)\Omega_{D-2}}\right)^{1/D-3}
\end{equation}
is the Schwarzschild radius; $M$ is the mass of the black hole and $d\Omega_{D-2}^2$ is the metric on the unit $D-2$-dimensional sphere. The area of the unit sphere is denoted by $\Omega_{D-2}$. The horizon of the black hole is at $r=R_S$. We consider the case for which the mass of the black hole is much greater than the mass of the particle, $M\gg m$, so that the backreaction on the Schwarzschild geometry is small.

Recall that in the Newtonian limit, the $00$-component of Einstein's field equation reduces to the Poisson equation for the Newtonian potential $\phi$ 
\begin{equation}
    {\nabla}^{2}\phi = \frac{8 \pi G \left(D-3\right)}{D-2}{\cal{\delta}}
\end{equation}
with ${\cal{\delta}}$ being the mass density.
Thus the Newtonian potential of a point particle of mass $m$ at position ${\bf r}_0$ is normalized in terms of the $D$-dimensional gravitational constant $G$ to be
\begin{equation}\label{eq:NGP}
    \phi = -\frac{8 \pi Gm}{\left( D-2 \right) \Omega_{D-2}} \frac{1}{\mid {\bf r} - {\bf r}_{0} \mid ^{D-3}}
\end{equation}

The black hole entropy is given by one quarter of the horizon area in Planck units
\begin{equation}
S= \frac{A}{4G}=\frac{R_S^{D-2}\Omega_{D-2}}{4G}
\end{equation}
and the Hawking temperature is 
\begin{equation}
T_H=\frac{D-3}{4\pi R_S}
\end{equation}
Our goal is to determine how these thermodynamic quantities get modified due to the presence of a perturbing mass near the horizon.  

In order to probe the near horizon region, we introduce a new coordinate $\rho$ that measures proper distance from the horizon:
\begin{equation}
\rho = \int_{R_S}^r {dr^{\prime} \over \sqrt{ f\left(r^{\prime}\right) } }
\end{equation}
%This is given explicitly by
%\begin{equation}
%\rho = \sqrt{r\left(r-2GM\right)} + 2GM \sinh^{-1}\sqrt{{r\over 2GM}-1}
%\end{equation}
In terms of $\rho$, the Schwarzschild metric is written as 
\begin{equation}
ds^2 = -f\left(r\left(\rho\right)\right)dt^2 + d\rho^2 + r^2(\rho)d\Omega^2
\end{equation}
A static observer at $r=\bar{r}$ is at distance $\rho(\bar{r})=\bar{\rho}$ from the horizon and measures a temperature
\begin{equation}
\overline T= {T_H \over \sqrt{f\left(\bar r\right)}} 
\end{equation}
Asymptotically this coincides with the Hawking temperature and increases as we move toward the horizon.
Of course the static observer accelerates relative to a freely falling observer and can explain the motion of a
freely falling particle in terms of a gravitational force.  

Next we set
\begin{equation}
{r - R_S \over R_S} = \epsilon, \,\,\,\, r=R_S(1+\epsilon)
\end{equation}
We take the black hole mass $M$ to be large, scaling $r$ as above, and focus in the small $\epsilon$ region.
In the large mass limit, the curvature invariants in this near horizon region become small, and we can approximate it as flat:
\begin{equation}
R_{\mu\nu\kappa\lambda}R^{\mu\nu\kappa\lambda} = {(D-1)(D-2)^2(D-3) \over {R_S}^4}\, \left(\frac{R_S}{r}\right)^{2(D-1)}
\end{equation}
So at the horizon
\begin{equation}
R_{\mu\nu\kappa\lambda}R^{\mu\nu\kappa\lambda} \sim {1 \over R_S^4}
\end{equation}
scaling to zero in the large mass limit.

Indeed when $\epsilon$ is small, $f\left( r \right)$ and $\rho$ admit the following expansions
$$
f\left(r\right)=\left[(D-3)\epsilon-\frac{(D-3)(D-2)}{2}\epsilon^2+\dots\right]
$$
\begin{equation}
\rho=\frac{2R_S}{\sqrt{D-3}} \left(\sqrt{\epsilon}+ \dots\right)
\end{equation}
where in the last expression the ellipses are of order $\epsilon^{3/2}$.
So
\begin{equation}
\epsilon= \left({\sqrt{D-3}\,\rho\over 2R_S}\right)^2 + \dots
\end{equation}
The metric becomes
\begin{equation}
ds^2 = -(D-3)\epsilon(\rho)\left[1-\frac{(D-2)}{2}\epsilon(\rho)+\dots\right]dt^2 + d\rho^2 +R_S^2\left(1+2\epsilon(\rho)+\dots\right)d\Omega_{D-2}^2
\end{equation}
In the region where $\epsilon \ll 1$ $\left(\rho \ll 2R_S/\sqrt{D-3}\right)$, the Schwarzschild metric can be approximated by
\begin{equation}
ds^2 \simeq -(D-3)\epsilon(\rho) dt^2 + d\rho^2 +R_S^2d\Omega_{D-2}^2=-\rho^2 \left({(D-3) \over 2R_S}\, dt\right)^2 + d\rho^2 +R_S^2d\Omega_{D-2}^2
\end{equation}
In the limit of large $M$, we may further approximate a sufficiently small patch of the sphere, e.g. around the positive $x_d$-axis, as flat \cite{Susskindb}. Note that this
patch becomes arbitrarily large in the limit $R_S \to \infty$.
So we end up with the $D$-dimensional flat Rindler metric
\begin{equation}
ds^2\simeq -\rho^2 d\omega^2 + d\rho^2 +\sum_{i=1}^{d-1}(dx_i)^2 
\end{equation}
where $\omega = (D-3)\, t/2R_S$ is a dimensionless time variable.

The horizon at $\rho =0$ becomes a planar Rindler horizon. However in computing the entropy shift,
it will be necessary
to take into account the compactness of the horizon and the finiteness of its area \footnote{The entropy density of the horizon is always maximal, given by one quarter in Planck units.}.
As long as $\bar \rho \ll 2R_S/\sqrt{D-3}$, the static observer lies in the flat region and records a temperature inversely proportional to the distance from the horizon: 
\begin{equation}
\overline T \simeq \frac{{T_H}}{\sqrt{(D-3)\bar \epsilon}} = {D-3 \over 4\pi R_S \sqrt{(D-3)\bar \epsilon}} \simeq {1 \over 2\pi \bar \rho}
\end{equation}
We recognize this as the Unruh temperature \cite{Unruh} associated with a uniformly accelerating observer.
The proper time associated with the clock of the Rindler observer at $\bar\rho$ is
\begin{equation}
d\tau_R=\bar \rho\, d\omega
\end{equation}

We define the coordinates 
\begin{equation}
t_M=\rho\sinh\omega, \;\;\;\; x_d=\rho\cosh\omega \label{Rindler}
\end{equation}
appropriate for a locally inertial (freely falling) observer.
The metric becomes the $D$-dimensional Minkowski metric:
\begin{equation}
ds^2 \simeq -(dt_M)^2 + (dx_1)^2 + \dots + (dx_d)^2 
\end{equation}
The trajectory $\rho = \bar \rho$ is equivalent with the hyperbola $(x_d)^2 - (t_M)^2 = \bar{\rho}^2$. The Rindler observer accelerates relative to an inertial observer. When $t_M=0$, the relative velocity is zero and the acceleration is $1/\bar \rho$. The Unruh temperature is $1/2\pi$ times this acceleration.

%%%%%%%%%%%%%%%%%%%%%%%%%%%%%%%%%%%%%%%%%%%%%%%%%%%%%%%%%%%%%%%%%%
\subsection{Test particle motion in the near horizon geometry}
%%%%%%%%%%%%%%%%%%%%%%%%%%%%%%%%%%%%%%%%%%%%%%%%%%%%%%%%%%%%%%%%%%%%
Consider a probe particle in the Schwarzschild geometry, moving along a radial geodesic.
Assume that the particle is instantaneously at rest at $r=r_0$ (along the positive $x_d$ axis).
The proper velocity and acceleration satisfy
\begin{equation}
\left({dr \over d\tau}\right)^2 = R_S^{D-3}\left({1\over r^{D-3}} -{1\over r_0^{D-3}}\right)\label{pmotion1}
\end{equation}
\begin{equation}
{d^2r \over d\tau^2}=-{ (D-3)R_S^{D-3} \over 2\, r^{D-2}} \label{pmotion2}
\end{equation}
\begin{equation}
{d^2r \over dt^2}= -{(D-3)R_S^{D-3} \over 2\, r^{D-2}}\left[1 - \left({R_S \over r}\right)^{D-3}\right] + {3(D-3)R_S^{D-3} \over 2\, r^{D-2}}\left({dr \over dt}\right)^2 {1 \over 1 -\left({R_S / r}\right)^{D-3} }
\label{pmotion3}
\end{equation}

Assume that $\epsilon_0 \ll 1$ but fixed; that is, the particle starts its motion in the near horizon region.
\ref{pmotion3} becomes
\begin{equation}
{d^2\rho \over d\omega^2}= - \rho + \dots \label{pmotion4}
\end{equation}
where the ellipses stand for velocity dependent terms, vanishing at the initial moment. This last equation can be obtained by studying
the motion directly in Rindler space. Indeed, from the point of view of the freely falling (Minkowski) observer,
the particle stays still at $x_d=\rho_0$. Then $\rho = \rho_0 / \cosh\omega$, producing \ref{pmotion4}.

Now suppose that the initial position of the particle coincides with the position of the observer at
$\bar \rho$ ($\rho_0 = \bar \rho$).
Relative to this observer, the particle's {\it initial} acceleration is
\begin{equation}
{d^2\rho \over d\tau_R^2}= -{1 \over \bar \rho}  \label{pmotion5}
\end{equation}
Thus, he concludes that an initial ($\omega =0$) attractive force acts on the particle, given by
\begin{equation}
F= -{m \over \bar \rho} \label{Force}
\end{equation}
where $m$ is the mass of the particle. We would like to interpret this force as an entropic force.

\subsection{Perturbing the near-horizon geometry of a large mass black hole}

We treat the black hole - particle system as an entropic system. The particle will move in a direction to maximize entropy. Once it gets absorbed by the black hole, there is an increase in the horizon radius and area, and the entropy increases. When the particle is at some distance from the horizon, there must be an entropic force, reproducing the gravitational attraction to the particle.
To this end, we study the perturbation of the near horizon geometry due to the presence of a small mass $m$, initially at rest on the positive $x_d$-axis,
at distance $\rho_0$ from the black hole horizon.

Having approximated the near horizon region as flat, we can easily obtain the leading effect due to the backreaction to the small
mass\footnote{For a study of perturbations in the full Schwarzschild metric see \cite{Peters, Zerilli}.}. 
As we discuss in \ref{A1}, the leading order perturbed metric is 
\begin{equation}
ds^2\simeq -(1+2\phi)\, dt_M^2 + \left(1-{2\phi \over D-3} \right) \sum_{i=1}^d(dx_i)^2 \label{pm1}
\end{equation}  
where 
\begin{equation}
\phi=-\frac{8\pi G m}{(D-2)\Omega_{D-2}} \; \frac{1}{\left(\sum_{i=1}^{d-1}(x_i)^2 +(x_d-\rho_0)^2\right)^{(D-3)/2}}
\end{equation}
Transforming to Rindler coordinates, \ref{Rindler}, we get for $\omega \simeq 0$
\begin{equation}
ds^2\simeq -\rho^2 (1+2\phi) \, d\omega^2 + \left(1-{2\phi \over D-3} \right) d\rho^2 + \left(1-{2\phi \over D-3} \right)\sum_{i=1}^{d-1}(dx_i)^2 \label{pm2}
\end{equation}
%This initially flat Rindler metric is an approximation to the near horizon geometry, close to the positive $x_d$-axis. Replacing the transverse space w%ith a sphere with unperturbed radius $R_S$, we obtain 
%\begin{equation}
%ds^2\mid_{\omega \simeq 0}\,\,\simeq -\rho^2 (1+2\Phi) d\omega^2 + \left(1-{2 \over D-3}\Phi \right) d\rho^2 + \left(1-{2 \over D-3}\Phi \right)R_S^2 d\Omega_{D-2}^2
%\end{equation}
The Rindler horizon is still at $\rho=0$ but the proper distance to it gets shifted. In order to obtain the area shift and the entropic force, we first compactify the Rindler horizon to a sphere of (unperturbed) radius $R_S$ so as to regulate its area, and take the large mass limit at the end. Letting $\omega \simeq 0$, the perturbation on this spherical horizon is given by
\begin{equation}
\phi_h(\vartheta)=-\frac{8\pi G m}{(D-2)\Omega_{D-2}} \; \frac{1}{\left(R_S^2 +L_0^2-2L_0R_S\cos\vartheta\right)^{(D-3)/2}}
\end{equation}
where $\vartheta$ is the angle of a point on the horizon with the positive $x_d$-axis; $L_0=R_S + \rho_0 > R_S$, where $\rho_0$ is
the initial distance of the particle from the horizon.

Thus to leading order in $m$, the shift in the horizon area is  
\begin{equation}
\delta A = -{D-2 \over D-3} \int_{S^{D-2}} \phi_h(\vartheta)\,R_S^{D-2}d\Omega_{D-2} \label{areashift}
\end{equation}
where we integrate the perturbation over the horizon sphere (of radius $R_S$). As we review in \ref{A2},
the integral can be evaluated exactly in terms of the potential at ``the center of the sphere'',
using the mean value theorem in $d=D-1$ spatial dimensions. The shift in the horizon area is 
\begin{equation}
\delta A = -{D-2 \over D-3}\Omega_{D-2}R_S^{D-2}\phi_C= \frac{8\pi G m R_S^{D-2}}{(D-3)\left(R_S+\rho_0\right)^{D-3}} 
\end{equation}

\begin{wrapfigure}{r}{0.5\textwidth}
\includegraphics[width=0.5\textwidth]{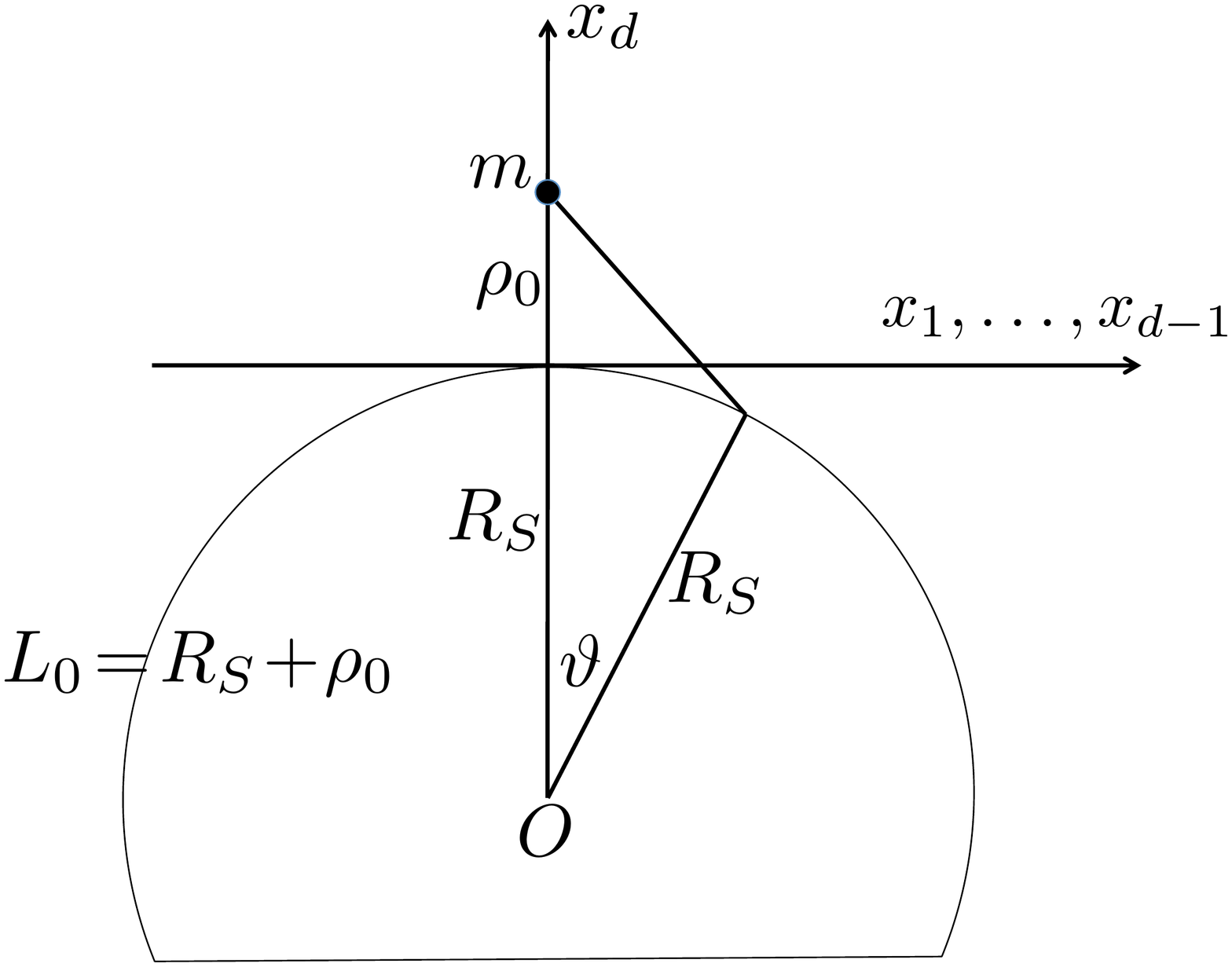}
\caption{The initial position of the test mass m. O is the center of the spherical horizon of radius $R_S$.}
\label{graph}
\end{wrapfigure}

Differentiating the expression above, and letting $R_S \to \infty$, gives
\begin{equation}
{d(\delta A) \over d \rho_0} = -8\pi Gm \label{areashift1}
\end{equation}
irrespectively of the number of spacetime dimensions $D$.

The shift in the horizon area amounts to a change in entropy. The corresponding entropy gradient in the large mass limit is
\begin{equation}
{dS \over d\rho_0}= {1 \over 4G} {d(\delta A) \over  d \rho_0} = - 2\pi m
\end{equation}
yielding an entropic force
\begin{equation}
F_{\rm entropic}= \overline T \, {d S \over d\bar \rho}= - {m \over  \bar \rho}
\end{equation}
This is in agreement with \ref{Force}.

\subsection{Considering charged black holes and spacetime rotation}
%%%%%%%%%%

Next we consider the case of an electrically charged test particle interacting with a large charged black hole in $D=4$. We would like to check if the gravitational force on the particle can be interpreted as an entropic force. In addition, the particle interacts with the electric field produced by the black hole\footnote{Notice that only the gravitational force on the charged particle should be obtained as an entropic force and not the net or the electric component of the force.}. So if Verlinde's theory is correct, the motion of the particle should be explainable via entropic and electromagnetic forces. We denote the charge and the mass of the test particle by $q$ and $m$, which we take to be sufficiently small.

The black hole geometry is described by the Reissner-Nordstrom metric, given by
\begin{equation}
ds^2 = -f\left( r \right)dt^2 + \frac{dr^2}{f\left( r \right)} + r^2d\Omega_{2}^2,\,\,\,\,\,f(r)=1-\frac{2GM}{r} + \frac{Q^2G}{r^2}\label{RSMetric}
\end{equation}  
where $Q$ is the charge of the black hole and $M$ its mass. The electric field is radial given by $E_r=Q/r^2$. To avoid a naked singularity we take $M^2G \ge Q^2$, and so
\begin{equation}
f(r) = \frac{(r-r_+)(r-r_-)}{r^2}  
\end{equation}  
with $r_{\pm} = MG\left(1 \pm \sqrt{1-Q^2/M^2G}\right)$. Thus the black hole has two horizons with radii $r_{\pm}$. The radius of the outer horizon is $r_+$. The associated entropy is given by $S=\pi r_+^2/G$ and the Hawking temperature by $T_H = \Delta/4\pi r_+^2$, where $\Delta = r_+ - r_-$. This temperature vanishes for an extremal black hole, for which $M^2G = Q^2$ and $r_+ = r_-$ -- see e.g. \cite{Susskindb} for more details. 

It is a rather challenging problem to obtain the backreaction on the black hole geometry (and calculate the shift in the horizon area) for generic motions of the particle, requiring one to incorporate relativistic effects, including the radiation emitted by the particle. In this work, we restrict to the case of a particle of sufficiently small mass and charge, instantaneously at rest in the near horizon region of a large, non-extremal black hole for which $M^2G > Q^2$. More specifically, we take the mass $M$ and the charge $Q$ of the black hole to be arbitrarily large, keeping the ratio $M^2G/Q^2$ fixed (and greater than unity). In this limit, the near horizon region, $r-r_+ \ll r_+$, becomes sufficiently weakly curved so as to obtain the backreaction\footnote{ The case of an extremal black hole, $M^2G = Q^2$, for which the near horizon geometry is $AdS_2 \times S_2$  \cite{Strominger} will be considered in future work.}.

Indeed in the near horizon region, the proper distance from the horizon is given by
\begin{equation}
\rho \simeq \frac{2 r_+ (r-r_+)^{1/2}}{\Delta^{1/2}}
\end{equation}
and the metric can be approximated by
\begin{equation}
ds^2 \simeq -\frac{\Delta^2\rho^2}{4 r_+^4}dt^2 + d\rho^2 + r_+^2 d\Omega_2^2
\end{equation}  
Furthermore, since we take $r_+$ to be very large, we approximate a sufficiently small patch of the sphere, around the positive $z$-axis, as flat. So we end with the four-dimensional flat Rindler metric:
\begin{equation}
ds^2 \simeq -\rho^2 d\omega^2 + d\rho^2 + dx^2 + dy^2
\end{equation}
where $\omega = \Delta t/2 r_+^2$. A static observer at small distance $\bar \rho \ll r_+$ from the horizon (along the positive $z$-axis) measures a temperature
\begin{equation}
\bar T = \frac{T_H}{\sqrt{f(\bar \rho)}}\simeq \frac{2 r_+^2 T_H}{\Delta \bar \rho}= \frac{1}{2 \pi \bar \rho}
\end{equation}  
Notice also that the electric field in this region becomes arbitrarily small in the limit: $E_r\simeq Q/r_+^2 \simeq Q/M^2G^2$.

Let the charged particle be instantaneously at rest at a distance $\bar \rho$ from the horizon along the positive $z$-axis. It is easy to see that the initial acceleration of the particle with respect to the static observer at $\bar \rho$ is given by
\begin{equation}
m {d^2\rho \over d\tau_R^2}= -{m \over \bar \rho} + q E_r \label{eomcharge}
\end{equation}
where $\tau_R=\bar \rho \omega$ is the proper time associated with the clock of the static observer at $\bar \rho$. We would like to check if the first term on the RHS can be interpreted as an entropic force. The second term is the electric force on the particle due to the interaction with the electric field of the black hole.

Since the near horizon region is approximately flat in the large $M, \, Q$ limit, we can obtain the backreaction on the geometry due to the presence of the charged particle. Keeping terms only linear in $m$ and $q$, the perturbed metric is given by \ref{pm1} in (locally) inertial coordinates and by \ref{pm2} in Rindler coordinates, as can be verified by examining the Reissner-Nordstrom metric in isotropic coordinates\footnote{Even if $q^2 > m^2G$, at a Compton wavelength away from the particle, the correction to the metric due to the electric field of the particle is suppressed (by a factor of $q^2$) compared to the correction $-2Gm/r$.}. So following similar steps as in the previous section, we can verify that the gradient of the shift in the horizon area is given by \ref{areashift1}. Therefore, the corresponding entropic force reproduces the first term of \ref{eomcharge}.

Finally it seems to be an interesting task to include the effect of spacetime rotation into the above discussions.
We take the line element of the Kerr spacetime in four spacetime dimensions~\cite{Carter:1968ks}
\begin{eqnarray}
ds^2=-\frac{\Delta_r-a^2\sin^2\theta }{\rho^2}dt^2 -\frac{2a \sin^2 \theta }{\rho^2 } \left( r^2+a^2 -\Delta_r\right)dt d\phi \nonumber\\ + \frac{\sin^2 \theta }{\rho^2 } \left( (r^2 +a^2)^2 -\Delta_r a^2 \sin^2\theta\right)d\phi^2 + \frac{\rho^2}{\Delta_r}dr^2 + {\rho^2}d\theta^2
\label{kds1}
\end{eqnarray}
where
$$\Delta_r = (r^2 +a^2) -2MGr,  \quad \rho^2 =r^2 +a^2 \cos^2 \theta $$
The parameter $a$ is called the rotation parameter. The black hole event horizon radius, $r_H$, is given by the largest positive root of $\Delta_r=0$. 

We could not find an appropriate generalization of the mean value theorem and \ref{potential} for \ref{kds1} for generic values of the rotation parameter. This is chiefly due to the lack of $SO(3)$ invariance of the spacetime. Instead, we wish to present a discussion on a weakly rotating version of the Kerr spacetime, keeping in terms only linear in $a$: 
\begin{equation}
ds^2 \approx - f(r) \, dt^2+f^{-1}(r)\, dr^2 +r^2 d\Omega^2 - 2a\sin^2 \theta \left(\frac{2MG}{r} \right)dt \,d\phi \label{kds2'}
\end{equation}
where $f(r)=1-2MG/r$.
%The term containing $H_0^2$ in the off-diagonal metric function cannot represent a true spacetime rotation but only the angular speed of the coordinate system we are using. Defining a new azimuthal coordinate by, $d\phi= d\tilde{\phi}+ aH_0^2 dt $, we get
%
%\begin{eqnarray}
%ds^2 \approx - f(r)dt^2+f^{-1}(r)dr^2 +r^2 d\Omega^2 -  \frac{4MGa \sin^2 \theta}{r} dt d\tilde{\phi}\, +{\cal O}(a^2)
%\label{kds2'}
%\end{eqnarray}
%
The surface gravity of the black hole event horizon is given by 
$$\kappa_H=\frac{f'(r=r_H)}{2}  $$
where the prime denotes differentiation with respect to the radial coordinate. 
The angular speed $=-g_{t{\phi}}/g_{{\phi}{\phi}}$ at the horizon is given by
$$\Omega_{H}=\frac{2MGa}{r_H^3}$$
Near the black hole event horizon, we make further coordinate transformation, $d{\phi}= d\bar{\phi} +\Omega_H dt$, e.g.~\cite{Cognola:1997dv}, to cast the metric in the diagonal form
\begin{eqnarray}
ds^2 \vert_{r\to r_H} \approx - f(r)\, dt^2+f^{-1}(r)\, dr^2 +r^2 \left(d\theta^2 + \sin^2 \theta \, d\bar{\phi}^2\right)  +{\cal O}(a^2)
\label{kds2}
\end{eqnarray}
which is analogous to \ref{SchwarzschildMetric} with $D=4$. We next follow similar steps described in the previous subsections to reach the same conclusions.

%When we are close to the cosmological event horizon, we use the coordinate transformation, $d \tilde{\phi}= d\phi' +\Omega_C dt$, in order to bring \ref{kds2} to a diagonal form.  We define the radial coordinate,
%$$\rho = \int_r^{r_C} \frac{dr}{\sqrt {f(r)}} $$
%Expanding $f(r)$ near the cosmological event horizon, $f(r\to r_C) \approx 2\kappa_C (r-r_C)$, we get
%$$\rho = -2 \sqrt{\frac{r-r_C}{2 \kappa_C}}$$ 
%Thus $\rho=0$ on the cosmological event horizon and since $\kappa_C<0$, it is negative inside. Consequently, the right hand side of \ref{Force} would have now opposite sign compared to the black hole event horizon, indicating repulsive effects. 

%The near cosmological horizon Rindler metric is formally similar to that of the black hole,
%$$ds^2 = -\rho^2 d\omega^2+d\rho^2 + r_C^2 d\Omega^2$$
%where $\omega= |\kappa_C| t $ is the dimensionless time variable. The check of entropic nature of the repulsive  force follows the same steps as \ref{s2}. 
%In \ref{areashift} $R_S$ has to be replaced with $r_C$ and also since we are inside the cosmological event horizon, $\rho_0$ has to be understood as negative.  Nevertheless, we obtain the force equation formally similar to \ref{Force} (with ${\bar \rho}<0$), verifying the entropic nature of the repulsive force.\\

Despite these agreements in the limit of large black hole mass, it is not clear to us that entropic forces can account for the gravitational forces and motion in generic systems with charges and/or angular momentum. In fact by examining simpler, spherically symmetric neutral systems in the next section, we argue that this is not the case.

%%%%%%%%%%%%%%%%%%%%%%%%%%%%%%%%%%%%%%%%%%%%%%%%%%%%%%%%%%%%%%%%%%%%%%%%%%%%%%%%%%

%%%%%%%%%%%%%%%%%%%%%%%%%%%%%%%%%%%%%%%%%%%%%%%%%%%%%%%%%%%%%%%%%%%%%%%%%%%%%%%%%%
\section{General spherically symmetric distribution}\label{s3}
%%%%%%%%%%%%%%%%%%%%%%%%%%%%%%%%%%%%%%%%%%%%%%%%%%
\noindent
In this section we generalize the computation to the case of a test mass moving in the field of an arbitrary spherically symmetric mass
distribution of finite radius (not necessarily comprising a black hole). Away from the distribution,
the geometry is described by the Schwarzschild metric, \ref{SchwarzschildMetric}, where $M$ is the total mass of the distribution.

It will be more convenient to work in isotropic coordinates. To this end we set 
\begin{equation}
r=R\left[1 + \frac{1}{4}\left(\frac{R_S}{R}\right)^{D-3}\right]^{2/(D-3)} \label{rR}
\end{equation}  
and the metric becomes \cite{Gao}
\begin{equation}
ds^2=-\left[\frac{1-\frac{1}{4}\left(\frac{R_S}{R}\right)^{D-3}}{1+\frac{1}{4}\left(\frac{R_S}{R}\right)^{D-3}}\right]^2dt^2+\left[1+\frac{1}{4}\left(\frac{R_S}{R}\right)^{D-3}\right]^{4/(D-3)}\left(dR^2 + R^2d\Omega_{D-2}^2\right)
\end{equation}  
When $R \gg R_S$, the metric can be approximated by 
\begin{equation}
ds^2\simeq-\left[1-\left(\frac{R_S}{R}\right)^{D-3}\right]dt^2+\left[1+\frac{1}{D-3}\left(\frac{R_S}{R}\right)^{D-3}\right]\left(dR^2 + R^2d\Omega_{D-2}^2\right) \label{IsotropicMetric}
\end{equation}  
Introducing Cartesian coordinates such that $\sum_{i=1}^d x_i^2=R^2$, this metric takes the form derived in \ref{A1}: 
\begin{equation}
ds^2\simeq -(1+2\Phi) \, dt^2 + \left(1-{2\Phi \over D-3} \right) \sum_{i=1}^d(dx_i)^2 \label{perturbedMetric}
\end{equation}  
where $\Phi=-8\pi GM/\left[(D-2)\Omega_{D-2}R^{D-3}\right]$ is the Newtonian potential associated with the mass distribution.

Next consider a static observer sufficiently far from the distribution, at fixed radial coordinate $R_0$ so that $R_0 \gg R_S$. Without loss of generality, we take the observer to lie on the positive $x_d$ axis. Define $x\equiv x_d-R_0$ and focus on a sufficiently small region $|x|, |x_i| \ll R_0$ around the location of this observer. There, the metric in \ref{IsotropicMetric} can be approximated by 
\begin{equation}
ds^2\simeq -\left(1+ 2g(R_0)\,x \right)dt^2 +dx^2 + \sum_{i=1}^{d-1}(dx_i)^2 \label{FlatMetric}
\end{equation}
where 
\begin{equation}
g(R_0)=\frac{8\pi G M\, (D-3)}{(D-2)\,\Omega_{D-2}}\; \frac{1}{R_0^{D-2}}
\end{equation}
is the acceleration due to gravity at radial distance $R_0$.
Notice that the metric \ref{FlatMetric} is flat.
Indeed the curvature invariant at $R_0$ is of order 
\begin{equation}
R_{\mu\nu\kappa\lambda}R^{\mu\nu\kappa\lambda}\mid_{R_0}\sim \frac{1}{R_0^4}\left(\frac{R_S}{R_0}\right)^{2(D-3)}
\end{equation}
so that a sufficiently small region of proper size $L \ll R_0$ around the static observer  can
be approximated to be flat.

Next consider the Rindler observer at $\rho = \bar \rho$, with uniform acceleration $1/\bar \rho$. Setting $\rho = \bar \rho + x$, we may expand the
Rindler metric for small $x$. We get
\begin{equation}
ds^2 \simeq -\bar \rho^2 \left( 1 + \frac{2}{\bar \rho}\, x\right) d\omega^2 + dx^2 + \sum_{i=1}^{d-1}(dx_i)^2=-\left( 1 + \frac{2}{\bar \rho}\, x\right)
d\tau_R^2 + dx^2 + \sum_{i=1}^{d-1}(dx_i)^2
\end{equation}
So locally we may identify the static observer with the Rindler observer if we set $\bar \rho = 1/ g(R_0)$. The static observer at $R_0$ records a temperature proportional to the acceleration due to gravity
\begin{equation}
T(R_0)={1 \over 2\pi \bar\rho}={g(R_0) \over 2\pi}\label{staticTemp}
\end{equation}
Construction of local thermal field theory using such local Unruh temperature can be seen in e.g.,~\cite{Buchholz:2006iv}.

The horizon corresponding to the Rindler observer is at distance
\begin{equation}
\bar \rho = \frac{1}{g(R_0)} \simeq R_0 \left(\frac{R_0}{R_S}\right)^{D-3}
\end{equation}
much outside the region which we can approximate as flat. So, we cannot apply the arguments of the previous
section on this horizon to compute the entropic force.

Instead, we use a spherical holographic screen of coordinate radius $R_0 \simeq r_0$ -- the Schwarzschild coordinate radius
$r_0$ is given by \ref{rR} -- that intersects the $x_d$ axis
at the location of the observer. This radius of the screen is sufficiently large to enclose the mass distribution.
According to the holographic principle \cite{tHooft, Susskind}, the interior region, and
everything that fits inside it, can be described in terms of a boundary theory on the screen. Furthermore, the number of microscopic
degrees of freedom of the boundary theory should scale with the area in Planck units. Indeed,
the maximal entropy allowed in the region is equal to the entropy of a Schwarzschild black hole that just fills the region \cite{Beke}.
This black hole has radius $r_0$, entropy equal to $A/4G$, and mass 
\begin{equation}
M_{\rm max}=\frac{(D-2)\Omega_{D-2}r_0^{D-3}}{16 \pi G}=\frac{(D-2)\left(\Omega_{D-2}A\right)^{(D-3)/(D-2)}}{16 \pi G} \label{Mmax}
\end{equation}
where $A$ is the proper area of the holographic spherical screen. The maximal entropy of any system is proportional to the number of fundamental degrees
of freedom needed to describe the system.

We suppose, as in \cite{Verlinde}, that in the underlying holographic description, the energy of the mass distribution is suitably partitioned among the microscopic degrees of freedom, so that the system becomes entropic with temperature given by \ref{staticTemp}. We will demonstrate in the following subsection that the equipartition principle and thermal equilibrium require the mass of the distribution to be an appreciable fraction of the maximal mass allowed in the region. Moreover, we argue below that the interpretation of gravity as an entropic force requires the entropy on the screen to be maximal.
If this is the case, the original mass distribution must be a black hole and the holographic screen the black hole horizon.        

We consider a freely falling particle of mass $m \ll M$, initially at rest at position $x_d=R_0+x_0$, $x_0>0$ on the $x_d$ axis. The particle lies in the region outside the screen. According to Verlinde, in the holographic description, the particle perturbs the system and the entropy changes as a function of the location of the particle. He postulates that when the particle is sufficiently close to the screen, the entropy gradient is proportional to the mass of the particle \cite{Verlinde}:
\begin{equation}
{dS \over dx_0}=-2\pi m \label{Entropyshift}
\end{equation}  
Then, indeed, the entropic force
\begin{equation}
F_{\rm entropic}= T(R_0) {d S \over dx_0}= - m g(R_0)
\end{equation}
coincides with the gravitational force on the particle at $R_0$.

But, let us take a closer look at the $x_0 \to 0$ limit. Let us, in particular, compute the shift in the area of the screen
due to the backreaction of the particle on the background geometry.
Since the screen radius $R_0 \gg R_S$ and the particle mass is taken to be very small, the perturbed metric is of the form \ref{perturbedMetric}
\begin{equation}
ds_{\rm perturbed}^2\simeq -(1+2\Phi^{\prime})\, dt^2 + \left(1-{2\Phi^{\prime} \over D-3} \right) \sum_{i=1}^d(dx_i)^2 
\end{equation}  
with $\Phi^{\prime}=\Phi + \phi_m$ being the total Newtonian potential,
including the contribution of the particle.
Using the metric above and the mean value theorem, we obtain the shift in the area of the holographic screen
\begin{equation}
\delta A = -{D-2 \over D-3} \int_{S^{D-2}} \phi_m(\vartheta)\,R_0^{D-2}d\Omega_{D-2} = \frac{8\pi G m\, R_0^{D-2}}{(D-3)\left(R_0+x_0\right)^{D-3}} 
\end{equation}
Differentiating with respect to $x_0$ and taking the $x_0 \to 0$ limit, gives
\begin{equation}
{d(\delta A) \over d x_0} = -8\pi Gm
\end{equation}
Therefore the entropy gradient \ref{Entropyshift} is precisely given by one quarter of the area gradient in Planck units:
\begin{equation}
 {1 \over 4G} {d(\delta A) \over  d x_0} = {dS \over dx_0} \label{SArelation}
\end{equation}

However, as we argue in the next subsection, the principle of equipartition and thermal equilibrium are compatible with \ref{SArelation} only if the entropy on the holographic screen is maximal. If this is the case, the original distribution must be a large black hole that fills the interior region and the screen the black hole horizon. The valid computation of the entropic force in the black hole case has been presented in \ref{s2}. It is not clear to us how to interpret the gravitational force in more general cases as an entropic force. Of course it could be that one or more of our assumptions break down, with the entropy on the screen being small and \ref{SArelation} holding, but we would like to understand how. Another possibility is to relate the entropy shift necessary to reproduce the gravitational force with a change in the entanglement entropy between the interior and the exterior regions. The entanglement area contains a divergent piece that scales with the area of the boundary (in units of the cutoff). However it is not clear how to regularize the entanglement entropy, including the fluctuations of the background geometry, and get the required coefficient.

Let us also make some comments about the choice of holographic screens. In this work we consider (quasi) static spacetimes with a time-like Killing vector $\xi^{\alpha}$ (that would become null at the horizon of a black hole), and which can be foliated with space-like surfaces that satisfy the holographic bound on the entropy \cite{Bousso}. The norm of the Killing vector, $-\xi^{\alpha}\xi_{\alpha}$, represents the redshift factor that relates the proper time of a static clock at some point in the interior of spacetime to that of a clock at infinity. As in \cite{Verlinde}, we take the screens to be surfaces of constant redshift. These would naturally include the horizon of a black hole and coincide with equipotential surfaces in regions of space with weak gravitational fields. Notice that the deformation of the equipotential surface due to the introduction of a test particle induces a higher order effect to the induced metric (and hence the area of the screen) with respect to the mass of the particle. In the cases of spherically symmetric spacetimes we are mostly interested in, the choice of screens is dictated by symmetry.

Equipotential surfaces play an important role in the original proposal of \cite{Verlinde}: the direction of the entropic force on a test particle, initially at rest at a small distance from the screen, is precisely given by the unit vector perpendicular to the screen. So this choice has a priori a chance to reproduce the gravitational force. Any other choice for screens (including dynamical ones)  should account for both the magnitude and direction of the force and also be compatible with black hole horizons.

It would be interesting to generalize the above results for non-spherically symmetric cases, to see how the notion of the entropic force would compare between surfaces with different topologies but otherwise with the same area (and hence the same maximal entropy). Clearly, in such cases
  %the use of a spherical polar coordinate system and the radial force will not be appropriate and perhaps
  one needs to develop a covariant and coordinate independent formalism to address the issue.

Finally for more general asymptotically flat spacetimes, one would need to foliate them with null hypersurfaces \cite{CJ} and make use of the covariant entropy bound \cite{Bousso} and other thermodynamics aspects of gravity. It is not clear to us however how to implement Verlinde's proposal in such situations.

%In such situations however that a holographic screen should be a  null hypersurface hiding information from an observer, so that the notion of entropy can be associated with it. In that sense the choice of the local Rindler horizon as the screen seems meaningful. If one instead is interested in the full general relativistic global spacetime, one must consider an appropriate null hypersurface in the asymptotic region as the holographic screen, in addition to the black hole event horizon, if any.  

%%%%%%%%%%%%%%%%%%%%%%%%%%%%%%%%%%%%%%%%%%%%%%%%%%%%%%%%%%%
\subsection{Statistical toy model}
%%%%%%%%%%%%%%%%%%%%%%%%%%%%%%%%%%%%%%%%%%%%%%%

Here we shall try to build a toy model in which \ref{SArelation} does not necessarily imply that the entropy on the holographic screen is maximal and the case studied above does not actually correspond to the black hole case studied in \ref{s2}.
According to the holographic principle, all configurations that fit in the interior region can be described in terms of a boundary theory living on the screen. The exterior region is empty space. Let us denote by $N$ the number of fundamental degrees of freedom in the boundary theory. The total energy in the system is given by the mass $M$ of the distribution, which cannot exceed $M_{\rm max}$ given by \ref{Mmax}. 

The energy $M$ will be distributed among the microscopic degrees of freedom $N$ according to some non-trivial distribution function. 
When $M$ is sufficiently large, we assume as in \cite{Verlinde} that the principle of equipartition holds to a good approximation. The resulting
temperature $T$ is proportional to the mean energy:
\begin{equation}
M=\frac{1}{2} N T
\end{equation}  
Requiring that the temperature is given by \ref{staticTemp},
the number of degrees of freedom $N$ scales with the area of the screen in Planck units, in accordance
with the holographic principle:
\begin{equation}
N \sim \frac{A(R_0)}{G}
\end{equation}  
%is the number of degrees of freedom and we set $k=(D-2)/(D-3)$.
Notice that the precise way the energy is divided among the microscopic degrees of freedom depends on the interactions and the details of the holographic
mapping, which we do not know.
The interactions may modify the precise relation between the energy and the temperature, but we expect the basic conclusions
regarding the realization of the entropic scenario to continue to hold.

Assuming thermal equilibrium, the first law gives (for fixed $N$)
\begin{equation}
dM= T dS = \frac{2M}{N} dS
\end{equation}
Integrating this equation, we obtain
\begin{equation}
 S = S_{\rm max} - \frac{N}{2}\ln\left({M_{\rm max} \over M}\right) = S_{\rm max} \left[ 1 -k \ln\left({M_{\rm max} \over M}\right)\right]=\frac{A}{4G}\,\left[ 1 -k \ln\left({M_{\rm max} \over M}\right)\right] \label{Sthermal}
\end{equation}
where $M_{\rm max}$ is the mass for which the entropy becomes maximal -- see \ref{Mmax} -- and $k=(D-2)/(D-3)$. In the last expression, we expressed the maximal entropy in terms of the area of the holographic screen. When the energy in the interior region becomes equal to $M_{\rm max}$, the system collapses to a black hole of radius $r_0$. Systems with greater energies do not fit in the interior region (and a screen of bigger area must be used).

For \ref{Sthermal}  to be valid, the mass $M$
cannot be arbitrarily small. In fact, thermal equilibrium and equipartition can be established when the mass of the distribution is an appreciable fraction of the maximal mass, namely
\begin{equation}
M \ge {M_{\rm max} \over  e^{1/k}}
\end{equation}
We expect that for smaller masses or temperatures, equipartition breaks down, with the temperature/energy relation depending strongly
on the details of the interactions of the boundary theory. 

Next consider the variation of the entropy, while keeping the mass $M < M_{\rm max}$ fixed.
Such a variation can arise from a shift in the area of the screen $\Delta A$, due to the shift of the location of the test particle. When $M = M_{\rm max}$,
$S= A/4G$ and $\Delta S = \Delta A/4G$. However when $M < M_{\rm max}$, we get   
\begin{equation}
\Delta S = \frac{\Delta A}{4G}\left[ 1 -k \ln\left({M_{\rm max} \over M}\right)\right] - k \frac{A}{4G}\frac{\Delta M_{\rm max}}{M_{\rm max}}
\end{equation}  
Using \ref{Mmax}, it is easy to see that
\begin{equation}
k\frac{\Delta M_{\rm max}}{M_{\rm max}}=\frac{\Delta A}{A}
\end{equation}  
in any spacetime dimension.
So 
\begin{equation}
\Delta S = -k \frac{\Delta A}{4G} \ln\left({M_{\rm max} \over M}\right)=-\frac{\Delta A}{4G}\left(1 - \frac{S}{S_{\rm max}}\right)  \neq  \frac{\Delta A}{4G} 
\end{equation}
Therefore if $M \neq M_{\rm max}$, it is not possible for the change of entropy to equal to $\Delta A /4G$.
Within our set of assumptions, this relation holds only for $M =  M_{\rm max}$ or $S = S_{\rm max}$.
If this is the case, the mass distribution already comprises a large black hole and was studied in \ref{s2}.

%%%%%%%%%%%%%%%%%%%%%%%%%%%%%%%%%%%%%%%%%%%%%%%%%%%%%%%%%%%%%%%%%%%%%%%%%%%%%%%%%%
\section{Discussion}\label{s4}
%%%%%%%%%%%%%%%%%%%%%%%%%%%%%%%%%%%%%%%%%%%%%%%%%%
\noindent

In this work we attempted to carry out some tests of the entropic gravity proposal in arbitrary spacetime dimensions ($D\ge 4$).
First we considered a large
Schwarzschild black hole interacting with a test particle, which is initially at rest at a small distance from the horizon.
The particle perturbs the near horizon geometry and induces a shift in the area of the horizon. We obtained the perturbation in the metric in
the limit of large black hole mass and at the moment the particle is instantaneously at rest.
As the mass of the black hole increases, the near horizon region becomes sufficiently weakly curved,
and can be well approximated with Rindler space. We then computed the shift in the horizon area, and hence the shift in the black hole entropy,
as a function of the particle's distance from the horizon. We found that the entropic force agrees
with the gravitational force on the particle (as seen by a static observer outside the black hole horizon),
irrespectively of the number of spacetime dimensions. The cases of a large charged black hole and a slowly rotating Kerr black hole were discussed next and the force on a test particle near the horizon was verified to satisfy the entropic force conjecture.

It would be interesting at this point to obtain the perturbed metric for more general particle motions and to incorporate
velocity dependent terms and relativistic effects, at least perturbatively. Then we could investigate if it is possible to reconstruct (perturbatively) the geodesic equations in terms of entropic forces and other thermodynamic quantities. Since entropic forces eventually cause irreversible changes in a system, it is not obvious if these can account for the most general particle motion in a gravitational background, including the case of a black hole background.
It would also be interesting to extend the study of the perturbations in the full Schwarzschild geometry, letting the mass of the black hole be large 
but finite and the distance of the particle from the horizon arbitrary.  

We then proceeded to generalize the computation to the case of an arbitrary spherically symmetric mass distribution, not necessarily
comprising a black hole. We chose a spherical holographic screen of sufficiently large radius to enclose the mass distribution, and associated to it an entropy and the Unruh temperature measured locally by a static observer. This temperature is dictated by the principle of equipartition of energy and the holographic scaling of the number of fundamental degrees of freedom on the screen. The entropy on the screen must be a small fraction of the maximal entropy allowed in the interior region.
We found that the entropy gradient needed to interpret the gravitational force on a nearby slowly moving test
particle as an entropic force is equal to one quarter of the area gradient in Planck units.
We argued that this result,
the principle of equipartition and thermal equilibrium imply that the entropy on the screen must be maximal,
given by one quarter of the screen area in Planck units.
So the original mass distribution must be a large black hole that fills the region inside the screen.
Within this set of assumptions, as usually stated, it is not clear
that the entropic gravity proposal can account for the gravitational interactions in the most general cases. This work, as we mentioned earlier, hopefully thus adds to the existing debates on the entropic gravity proposal, 
e.g.~\cite{Dai:2017qkz, Kobakhidze:2010mn}.

%%%%%%%%%%%%%%%%%%%%%%%%%%%%%%%%%%%%%%%%%%%%%%%%%%%%%%%%%%%%%%%%%%%%%%%%%%%%%%%%%%
\section*{Acknowledgements} NT wishes to thank the ITCP and the Department of Physics of the University of Crete where parts of this work were done for its hospitality. SB wishes to acknowledge S.~Chakraborty for discussions. SB's work is partially supported by the ISIRD grant 9-298/2017/IITRPR/704.
%%%%%%%%%%%%%%%%%%%%%%%%%%%%%%%%%%%%%%%%%%%%%%%%%%
\noindent

%%%%%%%%%%%%%%%%%%%%%%%%%%%%%%%%%%%%%%%%%%%%%%%%%%%%%%%%%%%%%%%%%%%%%%%%%%%%%%%%%

\bigskip
\appendix
\labelformat{section}{Appendix #1} 
%%%%%%%%%%%%%%%%%%%%%%%%%%%%%%%%%%%%%%%%%%%%%%%%%%%%%%%%%%%%%%%%%%%%%%%%%%%%%%%%%
%%%%%%%%
%%%%%%%%%%%%%%%%%%
%%%%%%%%%%%%%%%%%%%%%%%%%%%%%%%%%%%%%%%%%%%%%%%%%
\section{Metric Perturbation}\label{A1}
%%%%%%%%%%
Consider a system of non-relativistic particles moving in $D$-dimensional flat space. Assume that the particle masses are small
and the particle separations large, so that the gravitational forces between them are weak. As a first step,
we ignore the motions of the particles and treat the system as a static mass distribution.
Then we may obtain the corrections to the flat spacetime metric perturbatively.
We may also incorporate particle motion and obtain the relativistic corrections, employing the post-Newtonian approximation.
For the purposes of this work, it will be sufficient to obtain the leading correction to the metric, which is linear in the masses.
We write
\begin{equation}
g_{\mu\nu}=\eta_{\mu\nu}+h_{\mu\nu}
\end{equation}  
where $h_{\mu\nu}$ is the metric perturbation, linear in the masses.

Einstein's field equations can be written in terms of the energy-momentum tensor of the matter system as follows
\begin{equation}
R_{\mu\nu}=-8\,\pi\, G \, S_{\mu\nu}
\end{equation}
where
\begin{equation}
S_{\mu\nu} = T_{\mu\nu} - \frac{1}{D-2}T^{\lambda}_{\,\lambda}\, g_{\mu\nu}
\end{equation}
Ignoring the motion of the particles, the only non-vanishing component of the energy-momentum tensor is $T^{00}$,
which equals the density of rest mass $\delta$. Then 
\begin{equation}
S_{00}=\frac{D-3}{D-2}T^{00},\,\,\,\, S_{ij}=\frac{1}{D-2}\delta_{ij}T^{00}
\end{equation}
Imposing further the harmonic coordinate conditions
\begin{equation}
g^{\mu\nu} \,\Gamma^{\lambda}_{\mu\nu}=0
\end{equation}
and dropping corrections non-linear in the masses, the Ricci tensor simplifies to the following
\begin{equation}
R_{00}=\frac{1}{2}{\bf \nabla}^2 \,h_{00},\,\,\,\,\, R_{ij}=\frac{1}{2}{\bf \nabla}^2 \,h_{ij},\,\,\,\,\,R_{0i}=0
\end{equation}
Therefore, the $00$-component of Einstein's equation becomes to this order
\begin{equation}
{\bf \nabla}^2 \,h_{00}=-\frac{16\pi G(D-3)}{D-2}T^{00}
\end{equation}
with solution
\begin{equation}
h_{00}=-2\Phi
\end{equation}
where $\Phi$ is the Newtonian potential in $D$-dimensions satisfying
\begin{equation}
    {\bf \nabla}^{2}\Phi = \frac{8 \pi G \left(D-3\right)}{D-2}{\cal{\delta}}
\end{equation}
The spatial components give
\begin{equation}
{\bf \nabla}^2h_{ij}=-\frac{16\pi G}{D-2}\delta_{ij}T^{00}
\end{equation}
with solution
\begin{equation}
h_{ij}=-\frac{2}{D-3}\Phi\,\delta_{ij}
\end{equation}
The corrected metric is
\begin{equation}
ds^2\simeq -(1+2\Phi) dt_M^2 + \left(1-{2\Phi \over D-3} \right) \sum_{i=1}^d(dx_i)^2
\end{equation}  
%%%%%%%%%%%%%%%%%%%%%%%%%%%%%%%%%%%%%%%%%%%%%%%%%%%%%%%%%%%%%%%%%%%%%%%%%%%%%%%%%%%%%%%%%%%
\section{Mean Value Theorem}\label{A2}

The mean value theorem can be extended in $d$ spatial dimensions.
We start by applying Green's identities to the following integral over a $d$-dimensional ball $B_d$ of radius $R$, centered at the origin:
\begin{equation}\label{eq:MVT}
    \int_{B_d}  \left( \psi {\bf \nabla}^{2}\phi - \phi {\bf \nabla}^{2}\psi \right)d^{d} r = R^{d-1}\int_{S^{d-1}} \left( \psi {\bf \nabla}\phi - \phi {\bf \nabla}\psi \right) \cdot \hat{r}\, d\Omega_{d-1}
\end{equation}
We choose $\psi$ to be
\begin{equation}
    \psi \left( r \right) = -\frac{1}{\left( d-2 \right) \Omega_{d-1}} \left( \frac{1}{r^{d-2}} - \frac{1}{R^{d-2}} \right)
\end{equation}
which satisfies $\psi \left( R \right) = 0 $ and ${\bf \nabla}^{2}\psi = \delta^{d}\left({\bf r} \right)$, and $\phi$ to satisfy Laplace's equation
\begin{equation}
{\bf \nabla}^2 \phi = 0
\end{equation}
inside $B_d$. Then \ref{eq:MVT} reduces to the mean value theorem in $d$ spatial dimensions, namely
\begin{equation}
 \int_{S^{d-1}} \phi\, d\Omega_{d-1} = \Omega_{d-1}\, \phi_C
\end{equation}
where $\phi_{C}$ is the value of $\phi$ at the center of the sphere $S^{d-1}$. If ${\bf\nabla}^2\phi \neq 0$ inside $B_d$, then the integral gets an extra term
\begin{equation}
    \int_{S^{d-1}} \phi \, d\Omega_{d-1} = \Omega_{d-1}\phi_c -\Omega_{d-1} \int_{B_d}  \psi {\bf\nabla}^{2}\phi\, d^d {r}
\end{equation}

As an example, we apply the mean value theorem for the Newtonian potential of a point particle of mass $m$,
located at distance $L_0$ from the origin on the positive $x_d$ axis. In spherical coordinates, the potential is given by 
\begin{equation}\label{potential}
    \phi_m = -\frac{1}{\left( d-1 \right) \Omega_{d-1}} \frac{8\pi Gm}{\left( r^2  + L_0^2 - 2L_0r\cos{\vartheta} \right)^{\left(d-2\right)/2}} 
\end{equation}
where $\vartheta$ is the angle between the position vector ${\bf r}$ and the positive $x_d$ axis.
The mean value theorem gives
\begin{equation}
    \int_{S^{d-1}} \phi_m\, d\Omega_{d-1} = -\frac{1}{d-1}\,  \frac{8\pi Gm}{L_0^{d-2}}
\end{equation}
The center of the sphere is chosen to be at the origin and the radius $R$ is taken to be $R < L_0$.

The explicit calculation involves the integrals
$$
  K_n = \int_0^{\pi}  \frac{\sin^{n}{\vartheta}}{\left( R^2  + L_0^2 - 2L_0R\cos{\vartheta} \right)^{n/2}}\, d\vartheta = \frac{\sqrt{\pi}}{L_0^{n}} \, \frac{\Gamma\left((n+1)/2\right)}{\Gamma\left((n+2)/2\right)}
$$
\begin{equation} 
I_n = \int_0^{\pi} \sin^n{\vartheta}  d\vartheta = \sqrt{\pi}\; \frac{\Gamma\left((n+1)/2\right)}{\Gamma\left((n+2)/2\right)}
\end{equation}
The result is
$$
\int_{S^{d-1}}\phi_m\, d\Omega_{d-1} = -\frac{8 \pi GM}{\left( d-1 \right)\Omega_{d-1}}\, K_{d-2}\, \left( \prod_{n=1}^{d-3} I_n \right) 2\pi
$$
\begin{equation}
= -\frac{8 \pi Gm}{\left( d-1 \right)\Omega_{d-1}} \, \frac{1}{L_0^{d-2}} \, \frac{2\pi^{d/2}}{\Gamma\left(d/2\right)}
= -\frac{8 \pi Gm}{\left( d-1 \right)\Omega_{d-1}}\, \frac{1}{L_0^{d-2}}\, \Omega_{d-1}
= -\frac{8 \pi Gm}{d-1}  \,\frac{1}{L_0^{d-2}}
\end{equation}
in agreement with what the theorem predicts.

%%%%%%%%%%%%%%%%%%%%%%%%%%%%%%%%%%%%%%%%%%%%%%%%%%%%%%%%%%%%%%%%%%%%%%%%%%%%%%%%%%

\end{document}